\documentclass{emulateapj}
\usepackage{apjfonts,graphicx,hyperref,epsf}
\usepackage{graphicx}
\usepackage{subfigure}
\begin{document}
\shortauthors{Morningstar et al.}

\title{The spin of the Black Hole GS 1124-683:  Observation of a Retrograde Accretion Disk?}

\author{Warren~R~Morningstar,\altaffilmark{1}
        Jon~M.~Miller,\altaffilmark{1}
        Rubens~C.~Reis,\altaffilmark{1}
        Ken~Ebisawa\altaffilmark{2}}

\altaffiltext{1}{Department of Astronomy, University of Michigan, 500 Church Street, Ann Arbor, MI 48109-1042, wmorning@umich.edu, jonmm@umich.edu}

\altaffiltext{2}{Institute of Space and Astronautical Science (ISAS), Japan Aerospace Exploration Agency (JAXA), 3-1-1 Yoshino-dai, Chuo-ku, Sagamihara, Kanagawa 252-5210, Japan}


\label{firstpage}

\begin{abstract}

We re-examine archival {\it Ginga} data for the black hole binary system GS 
1124-683, obtained when the system was undergoing its 1991 outburst.  Our 
analysis estimates the dimensionless spin parameter $a_{*}=cJ/GM^{2}$ by 
fitting the X-ray continuum spectra obtained while the system was in the 
``Thermal Dominant'' state.  For likely values of mass and distance, we find 
the spin to be $a_{*}=-0.25_{-0.64}^{+0.05}$ (90\% confidence), implying that 
the disk is retrograde (i.e. rotating antiparallel to the spin axis of the 
black hole).  We note that this measurement would be better constrained if the
distance to the binary and the mass of the black hole were more accurately 
determined.  This result is unaffected by the model used to fit the hard 
component of the spectrum.  In order to be able to recover a prograde spin,
the mass of the black hole would need to be at least $15.25~M_{\odot}$, or the
distance would need to be less than $4.5~{\rm kpc}$, both of which disagree
with previous determinations of the black hole mass and distance.  If we allow
$f_{\rm col}$ to be free, we obtain no useful spin constraint.  We discuss 
our results in the context of recent spin measurements and implications for jet
production.

\end{abstract}

\section{Introduction}

Spinning black holes (BHs) are of fundamental importance to astrophysics, 
because they represent laboratories for the exploration of General Relativity. 
Spin is constrained by indirect measures involving the accretion disk.  A 
low-mass X-ray binary (LMXB) is an example of a binary in which the black hole
(or neutron star) is orbited by a small star, usually with a mass less than 
that of the Sun.  The star usually fills its Roche lobe and its outermost 
layers of gas are stripped off its surface by the immense gravity of the 
compact object, and form an accretion disk.

In periods of increased accretion activity, X-ray Novae can occur.  X-ray Novae
are generally transient phenomena, waiting on average 10-50 years between 
outbursts (Tanaka \& Shibazaki 1996).  Attempts to create a unified model of 
the disk evolution in these outbursts (Esin et al. 1997) have led to the 
description of the outbursts as the combination of a thin accretion disk and an
Advection Dominated Accretion Flow (ADAF).  Observationally, these two migrate
through several spectral states (see e.g. Reynolds \& Miller 2013).  The state 
important to this analysis is the 
Thermal Dominant State (TD state, formerly referred to as High State or 
High/Soft state), in which the disk extends all the way to the innermost stable
circular orbit (ISCO), and dominates the emission (Esin et al. 1997).  Since 
the ISCO is entirely dependant on the spin ($r_{\rm ISCO}=6r_{g}$ for a 
schwarzschild BH, $r_{\rm ISCO}=r_{g}$ for a maximally prograde BH, $r_{\rm 
ISCO}=9r_{g}$ for a maximally retrograde BH), measurements of the radius of the
ISCO can be used to determine the spin of the BH.  This is the idea behind the 
continuum fitting method, in which one fits a model of a thin accretion disk to
Thermal Dominant State spectra of a BH to estimate the ISCO, and thus infer the
spin (Zhang, Cui, \& Chen 1997, Schafee et al. 2006, McClintock et al. 2006, 
etc.).  Spin can also be measured by modeling the broadened Iron K-shell 
emission line that originates in the inner disk (e.g. Tanaka et al 1995, Miller
et al. 2002, Miller et al. 2004).

A common XSPEC model to describe the thin accretion disk is {\it kerrbb} 
(Arnaud 1996; Li et al. 2005).  This model stands out since it takes the 
spin as a parameter used to define the model spectrum.  It also includes 
relativistic effects such as limb-darkening or self irradiation of the disk. In
order for this method to be used to estimate spin, the mass, distance, and 
inclination angle of the disk must be known (Zhang, Cui, \& Chen 1997).  
Additionally, the spin measurement is dependent on the color correction factor
($f_{\rm col}=T_{\rm col}/T_{\rm eff}$).  Again, {\it kerrbb} is useful, since 
it accepts all of these as input model parameters.  Other new models, 
such as {\it simpl} (Steiner et al. 2009a) offer
an improved description of the hard component relative to a powerlaw.  The 
pairing of {\it kerrbb} and {\it simpl} have been 
used several times to measure spin (see for example Gou et al. 2009, Steiner et
al. 2010, etc).

\section{Source and Data Selection}

\begin{figure*}[htbp]
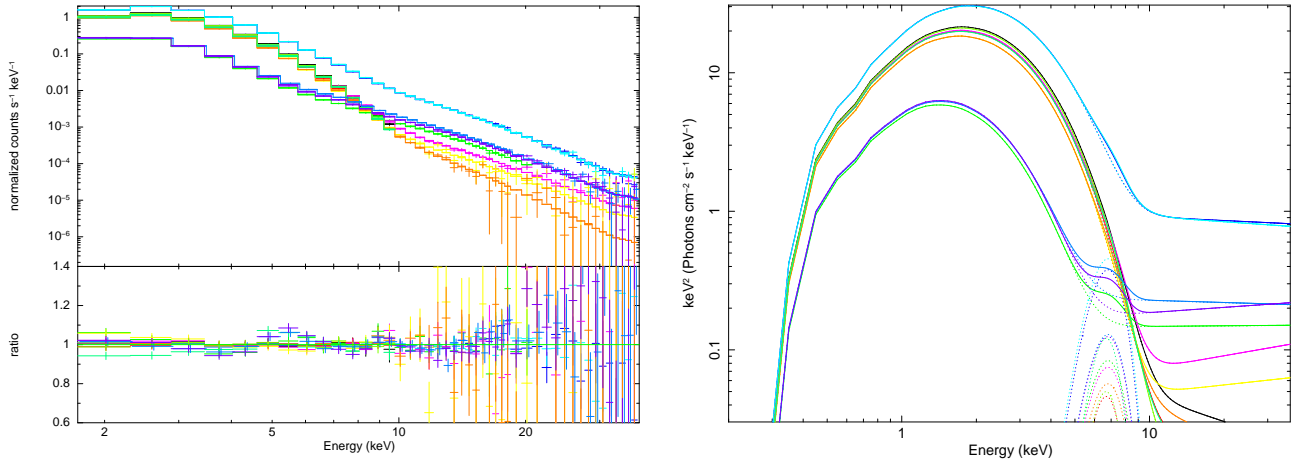

\begin{center}
\subfigure{\includegraphics[scale=0.35, angle =270]{f1.ps}}
\subfigure{\includegraphics[scale=0.35, angle =270]{f2.ps}}
\caption[b]{\footnotesize (Left) All spectra of GS 1124-683 used in our 
analysis, and the data to model ratios.  The large errors above 20 keV are as a
result of diminishing sensitivity at higher energy. (Right) Best fit model to 
our data.  In all cases the disk dominates the emission, and in several of the 
spectra, the hard component becomes very faint}
  \vspace{0cm}
  \label{fig:samplespectra}
\end{center}
\end{figure*}

GS 1124-683 (also called Nova Muscae 1991) is a LMXB that underwent an outburst
in 1991.  It was discovered in January 1991 by the All Sky monitors on both the
{\it Ginga} satelite, and the {\it Granat} satelite (Makino et al. 1991; Lund 
\& Brandt 1991; Kitamoto et al. 1992; Brandt et al. 1992). It flared up to a 
maximum flux of 8 Crab 
($1.92\times10^{-7}~{\rm ergs}~{\rm cm^{-2}}~{\rm s}^{-1}$) 
on January 15, and subsequently decayed exponentially with a timescale 
$\tau=30~{\rm days}$ (Ebisawa et al. 1994).  
It was studied using {\it Ginga} (Ebisawa et al. 1994) over the
course of several months, during which it migrated through all 5 of the typical
spectral states.

The BH mass, distance, and inclination  have been refined several times.  
Shahbaz et al. (1997) modeled the infrared light curve to deduce the BH mass, 
the mass of the secondary star, the binary separation, and the binary 
inclination.  They also inferred from these the distance to the BH using 
Bailey's relation (Bailey 1981).  In fact the distance to GS 1124-683 has been 
revised by several authors (Della Valle et al. 1991; Orosz et al. 1996; 
Shahbaz et al. 1997; Gelino et al. 2001), the most recent of which being Gelino
(2004, Hereafter G04), who refined the method from Gelino et al. (2001; the 
refined method is described in Gelino, Harrison, \& Orosz 2001) and 
found the distance to be $5.89\pm0.26~{\rm kpc}$, which falls within the range 
allowed by Orosz et al. (1996), but is better constrained.  The inclination 
angle measurement is much better agreed upon, with the value from G04 of 
$54^{\circ}\pm1^{\circ}.5$ agreeing with that from Shahbaz et al. 
(1997).  The mass is also fairly well agreed upon, with the measurement of 
$7.24 \pm 0.70~{\rm M_{\odot}}$ from G04 corresponding to those made previously
by Shahbaz et al. (1997), and those made by Orosz et al. (1996).  We used the 
best fit values from G04 for mass, distance, and inclination because they are 
newer and better-constrained than other determinations, and since they were 
found using infrared photometry, from which the disk and hotspot produce less 
contamination in the light curve.  It should also be 
noted that we assume the inclination angle of the inner disk to be the same as 
that of the binary, which is not necessarily the case (e.g. Maccarone 2002).  

The X-ray data we consider are those presented in Ebisawa et al. (1994).  For 
the continuum fitting method, we want to use spectra obtained when the source 
was in the TD state.  As per McClintock et al (2006), we selected disk 
luminosities less than 30\% of the Eddington limit, and restricted our 
observations to those in which the soft flux contributes at least 90\% of the 
total flux ($F_{\rm soft}/F_{\rm tot}\geq0.9$) based on the results reported
in Ebisawa et al. (1994).  Assuming a distance of $5.89~{\rm kpc}$ 
(G04), we find that the peak luminosity reached was 
$7.97\times10^{38}~{\rm ergs}~{\rm s^{-1}}$, which, assuming a black hole mass
of $7.24~{\rm M_{\odot}}$ (G04), is about $0.87~L_{\rm edd}$.  Assuming an
exponential decay with a timescale of 30 days (Ebisawa et al. 1994), we find 
that observations falling into our luminosity criterion begin 32 days after 
January 15 (Feburuary 16).  Observations falling into our hardness criteria 
began on February 16 as well, and ended on May 18, when the source transitioned
to the Low/Hard state.

Some spectra required additional consideration due to anomalous
behaviors they exhibited.  As noted in Ebisawa et al. (1994), observations 
occuring in late March and throughout April had a hard component that was too 
faint to be obseerved by the {\it Ginga} detectors.  For those spectra (March 
28, 29, 30, and April 2), we were required to ignore outside of the energy 
range 1.2-10 keV.  For the first of the May 17 spectra, the hard component 
became too faint to be observed at energies exceeding ~25 keV, so we ignored 
those energies in that spectrum.  We also ignored the April 19 observation 
altogether because it required an excessively low color correction factor in 
order for the spin to be consistent with the other observations (1.36), and 
because it had a $\chi^{2}_{\nu}$ value that was too high ($\sim3$) when
$f_{\rm col}$ was required to be within our allowed range (1.5-1.9, see 
Shimura \& Takahara 1995).  For all other observations, we examined over the 
entire reliable energy range for data obtained with {\it Ginga}; 1.2-37.0 keV 
(Ebisawa 1991).

\section{Analysis}
\label{sec:Analysis}

All analysis was performed in XSPEC version 12.8.0 (Arnaud 1996).  The model 
central to our analysis is {\it kerrbb} (Li et al. 2005), which models a thin
accretion disk around a kerr black hole.  {\it Kerrbb} is convolved with 
{\it simpl} (Steiner et al. 2009a), an empirical model for Comptonization. This
model provides a more physical description of the hard component (as opposed
to {\it powerlaw}), and yields fits of equal statistical quality.  It also 
has the virtue of simplicity compared to more rigorous models of 
Comptonization (such as {\it compTT} or {\it compBB}).

In addition, we included the effects of absorption by the interstellar medium,
{\it tbabs} (Wilms, Allen, \& McCray 2000).  We fixed the hydrogen column 
density to $1.5 \times 10^{21}{\rm cm^{-2}}$, which is the best-fit value found
for $N_{H}$ in Ebisawa et al. (1994).  We also found it necessary to add a 
Gaussian line with energy 6.5 keV, and with its width allowed to vary between 0
and 1 keV.  Relativistic lines did not improve the fit by a statistically 
significant margin.  Altogether, this model is shown in 
Figure~\ref{fig:samplespectra}.

We fixed the mass, distance, and inclination to 
the measurements given by G04, and fixed the norm of {\it 
kerrbb} to 1, as should be done when mass, distance, and inclination are fixed 
(Shafee et al. 2006, McClintock et al. 2006, etc.).  We did not include the
effects of limb-darkening.
We allowed the spin ($a_{*}$), effective mass accretion rate ($\dot{M}$), 
photon index ($\Gamma$), and fraction of the seed photons scattered 
($f_{\rm sc}$) to vary freely and unconstrained, and the color correction 
factor $f_{\rm col}$ to vary between 1.5 and 1.9 (Shimura and 
Takahara 1995).  We added a 2 percent systematic error to all energy bins to 
ensure acceptable fits, typical for analyses of {\it Ginga} data.

We fitted spectra individually at first.  For those fits that ignored large 
portions of the hard component, the spins were not very well constrained.  In 
order to place tighter constraints on the spin, we found it better
to jointly fit all spectra.  For joint fits, we required the spin, and spectral
hardening factor to be jointly determined, and allowed the rest to vary as 
before.

To examine the full allowable parameter space (since we do not 
have entirely precise measurements of mass and distance), we did a $3\times3$ 
grid search of mass and distance, fitting the spectra for each pairing, finding
the best fit parameters, and estimating their uncertainties.  The points on 
our grid correspond to the best fits of mass and distance from G04, and their 
upper and lower limits.  The uncertainty found here 
propagates into the uncertainty in our spin measurement, since it is entirely 
allowed that the BH could have any coupling of parameters within that grid.  
To globally cover this space, and to find the best-fit values of all 
parameters, we fitted with mass and distance required to be jointly determined,
but allowed to vary within this grid range.

\section{results}

\begin{table*}[tb]
\caption[t]{Spectral Fitting Results}
\label{tab:par}
\begin{center}
\begin{tabular}{llllllll}
\tableline \tableline
Time & $\Gamma$ & $f_{\rm sc}$ & $a_{*}$ & $\dot{M}$ & $f_{\rm col}$ & $\sigma$ & norm \\
(UT; 1991)  & ~ & ($10^{-3}$) & ($cJ/GM^{2})$ & ($10^{18}~{\rm g}~{\rm s}^{-1}$) & ~ & (keV) & ($10^{-3}~{\rm photons}~{\rm cm}^{-2}~{\rm s}^{-1}$) \\[1ex]
\tableline

2/20 ~ 23:31-23:36 & $2.08_{-0.03}^{+0.04}$ & $13.9^{+1.0}_{-0.7}$ & $-0.25^{+0.05}_{-0.64}$ & $8.0_{-0.2}^{+4.1}$ & $1.51^{+0.10}_{-0.01}$ & $1.0_{-0.07}^{+0}$ & $22\pm3$ \\[1ex]

2/21 ~ 0:23-0:29 &  $2.12^{+0.04}_{-0.03}$ & $14.8_{-0.8}^{+1.1}$ & $-0.25^{+0.05}_{-0.64}$ & $8.0_{-0.2}^{+4.1}$ & $1.51^{+0.10}_{-0.01}$ & $1.00^{+0}_{-0.04}$ & $26^{+4}_{-3}$  \\[1ex]

3/8 ~ 18:04-18:21 & $1.68^{+0.11}_{-0.09}$ & $0.9^{+0.2}_{-0.1}$ & $-0.25^{+0.05}_{-0.64}$ & $5.5^{+2.8}_{-0.1}$ & $1.51^{+0.10}_{-0.01}$ & $1.0_{-0.03}^{+0.0}$ & $4.1^{+1.0}_{-0.7}$  \\[1ex]

3/10 ~ 16:56-17;16 & $1.8\pm0.3$ & $0.8_{-0.3}^{+0.5}$ & $-0.25^{+0.05}_{-0.64}$ & $5.0_{-0.1}^{+2.6}$ & $1.51_{-0.01}^{+0.10}$ & $1.0_{-1.0}^{+0}$ & $0.7_{-0.7}^{+1.1}$   \\[1ex]

3/20 ~ 12:56-13:56 & $2.8^{+0.3}_{-0.2}$ & $2.5_{-0.7}^{+1.4}$ & $-0.25^{+0.05}_{-0.64}$ & $5.0_{-0.1}^{+2.6}$ & $1.51_{-0.01}^{+0.10}$ & $1.0_{-0.1}^{+0.0}$ & $3.1_{-0.7}^{+1.0}$\\[1ex]
 
3/28 ~ 9:37-9:42 & $2.6~(frozen)$ & $2.0^{+0.4}_{-0.3}$ & $-0.25^{+0.05}_{-0.64}$ & $5.8_{-0.1}^{+3.0}$ & $1.51_{-0.01}^{+0.10}$ & $0.4_{-0.4}^{+0.6}$ & $0.7_{-0.7}^{+1.5}$  \\[1ex]

3/29 ~ 5:54-6:05 & $2.6~(frozen)$ & $0.6\pm0.3$ & $-0.25^{+0.05}_{-0.64}$ & $5.7_{-0.1}^{+2.9}$ & $1.51_{-0.01}^{+0.10}$ & $0.8_{-0.3}^{+0.2}$ & $2^{+1.2}_{-0.8}$ \\[1ex]

3/30 ~ 8:36-8:53 & $2.6~(frozen)$ & $1.1^{+0.4}_{-0.2}$ & $-0.25^{+0.05}_{-0.64}$ & $5.4_{-0.1}^{+2.8}$ & $1.51^{+0.10}_{-0.01}$ & $0.8\pm0.1$ & $5.4_{-0.7}^{+1.1}$\\[1ex]

4/2 ~ 5:04-5:29 & $2.6~(frozen)$ & $1.1_{-0.2}^{+0.9}$ & $-0.25^{+0.05}_{-0.64}$ & $5.4_{-0.1}^{+2.8}$ & $1.51_{-0.01}^{+0.10}$ & $0.8\pm 0.2$ & $2.0_{-0.7}^{+1.1}$  \\[1ex]

5/17 ~ 3:12-3:19 & $1.98^{+0.08}_{-0.07}$ & $9_{-1}^{+2}$  & $-0.25^{+0.05}_{-0.64}$  & $1.73_{-0.03}^{+0.9}$ & $1.51_{-0.01}^{+0.10}$ & $1.00_{-0.07}^{+0.0}$ &  $4.8^{+0.5}_{-0.6}$  \\[1ex]

5/17 ~ 4:34-4:56 & $1.88_{-0.04}^{+0.05}$ & $9.0_{-0.7}^{1.1}$ & $-0.25_{-0.64}^{+0.05}$ & $1.85_{-0.03}^{+0.95}$ & $1.51_{-0.01}^{+0.10}$ & $1.00_{-0.05}^{+0.00}$ & $7.0_{-0.5}^{+0.4}$   \\[1ex]

5/17 ~ 7:49-8:09 & $2.05\pm0.04$ & $15\pm1$ & $-0.25^{+0.05}_{-0.64}$ & $1.81_{-0.03}^{+0.93}$ & $1.51_{-0.01}^{+0.10}$ & $1.00_{-0.02}^{+0.00}$ & $7.4\pm 0.5$  \\[1ex]

\tableline
\end{tabular}
\end{center}
\tablecomments{Results of joint spectral fits to the observations in our 
sample.  The $\chi^{2}$ value for the best fit is 439.57 (342 DOF).  }
\end{table*}

\begin{figure}[tb]
\includegraphics[scale=0.42]{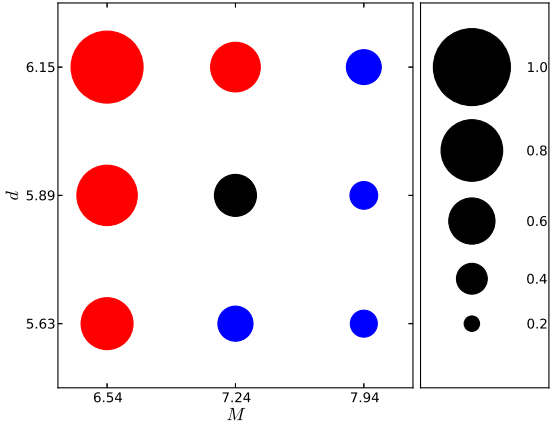}
\caption{\footnotesize Results of a Grid search through the mass/distance
parameter space.  The size of the circles is proportional to the magnitude of 
the spin $|a_{*}|$.  Circles filled blue have lower $\chi^{2}$ than the fit 
using the best-fit mass and distance in G04 
($439.0\leq\chi^{2}\leq441.5$), and the red circles have higher $\chi^{2}$ 
($443.0\leq\chi^{2}\leq447.0$).  All above fits have $\nu=344$.} 
  \vspace{0cm}  
  \label{fig:1124bubbleplot}
\end{figure}

\begin{figure}[tb]
\includegraphics[scale=0.42]{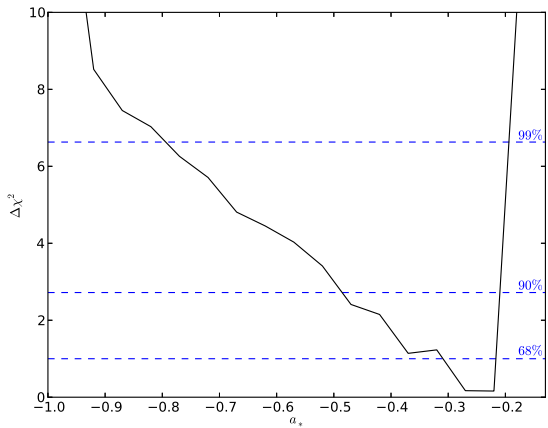}
\caption{\footnotesize Spin Contours for GS 1124-683 obtained while holding 
mass and distance within the parameter space used for the grid search, but 
allowing them to be variable, jointly-determined parameters.  Horizontal 
dotted lines are drawn at the 68\%, 90\%, and 99\% confidence levels.}
  \vspace{0cm}
  \label{fig:1124spincontours}
\end{figure}

Table~\ref{tab:par} shows the results of spectral fitting.  The best-fit value
of the spin is $a_{*}=-0.25^{+0.05}_{-0.64}$, implying that the spin is 
retrograde.   The lower bound here is very relaxed, since our analysis
allows uncertainty to propagate from uncertainties in the mass and distance
without taking account of the statistical preferences
inside of our allowed range.  Figure~\ref{fig:1124bubbleplot} shows the result 
of the grid search, which expresses the extent at which different pairings of 
mass and distance affect our measurement of the spin.  For all parameters, the
uncertainties expressed in Table~\ref{tab:par} reflect the upper and lower
limits estimated from our grid search.  The $\chi^{2}_{\nu}$ values for this 
analysis favor a smaller magnitude of the spin, smaller magnitudes of the 
distance, and larger masses for the BH than the best-fit values from G04.

To place a tighter constraint on $a_{*}$ while taking account of 
our uncertainties in the mass and distance, we allowed M and d to 
vary within our grid range but kept them jointly determined.  We did fits 
stepping through 20 evenly spaced values of $a_{*}$ between -0.97 and 0.03 
using {\it steppar}.  Figure~\ref{fig:1124spincontours} shows the result.  
Here the best fit value for the spin lies between -0.5 and
-0.2 with 90\% confidence, which is better-constrained than the value 
estimated from the grid search since it takes account of the behavior of
$\chi^{2}$ with respect to mass and distance rather than the grid search, which
treats each pairing as equally likely.  The spin here is greater than -1.0 at 
just over $3\sigma$, is less than -0.15 at the $6\sigma$ level, 
and is less than 0 at $>>8\sigma$ (the p-value for $a_{*}=0$ is 
$\sim10^{-102}$).  We choose to use the confidence range estimated from our 
grid search however, since it was calculated with M and d fixed, which is a 
necessity for finding spin by fitting the continuum.

As a check on our results, we decided to examine the behavior of $a_{*}$
as we changed certain other parameters, namely the color correction factor 
$f_{\rm col}$.  This is a useful check, since our value of $f_{\rm col}$ is 
close to the minimum of the allowed range.  We fixed its value to 1.7 and fit
all other parameters.  We find that raising the value of $f_{\rm col}$ even 
this high results in the spin immediately being pegged at $a_{*}=-1$, the 
theoretical limit.  This further solidifies our determination of $a_{*}$ as 
being retrograde.

\section{Discussion}

A retrograde spin is atypical in a black hole LMXB.  Nonetheless our 
measurement is consistent with previous attempts to measure the spin for GS 
1124-683.  Suleimanov et al (2008) constrained it to be $\leq 0.4$, and Zhang, 
Cui, and Chen (1997) estimated it to be nearly Schwarzschild, yet retrograde 
(-0.04).  Although our method is descendant from Zhang et al. (1997), ours 
yields a different measurement of the spin since it takes advantage of newer 
models developed ({\it kerrbb} \& {\it simpl}), which provide a more physical 
description of the spectrum, and allow for relativistic effects.  Recent works 
(Reis et al. 2013, Gou et al. 2010) have measured spins for SWIFT J1910.2-0546 
and A06200-00 that are either retrograde, or consistent with being retrograde, 
implying that such spins are less rare than previously thought.

Furthermore, as shown in Figure~\ref{fig:1124spincontours}, our upper limit to 
the spin is still retrograde.  If the spin is actually prograde one of two 
things must be true.  Either the distance must be 
considerably smaller than that given by G04, or the BH mass must be larger. To 
find out how much closer, or more massive it needs to be, we fixed either mass 
or distance to their best determination from G04, and slowly raised or lowered 
the other until their best fit value of $a_{*}$ was greater than 0.  When we 
varied the mass, and held the distance constant, we found that GS 1124-683 must
have $M_{BH}\geq15.25~M_{\odot}$.  This is in disagreement not only with 
G04, but also with Shahbaz et al. (1997), who constrained the BH 
mass to be less than $10.5~M_{\odot}$ at the 90\% confidence level using the 
maximum mass of the secondary star (Inferred from the spectral type).  This 
makes a mass of $15.25~M_{\odot}$ unlikely, so if the black hole is 
prograde, it is more likely that the distance to the binary is lower.  In fact,
in order for the spin to be prograde, the distance 
would have to be $d\leq4.5~{\rm kpc}$, still greater than the maximum allowed 
distance from Shahbaz et al. (1997).  If we take their entire range of mass and
distance into account however, and perform a grid search, we 
find that the spin found using their best determination of the mass and 
distance is still retrograde ($a_{*}=-0.16$ for ${\rm M}=5.8~{M_{\odot}}$ and 
${\rm d}=4~{\rm kpc}$), there is no lower bound on $a_{*}$, and the upper 
bound is $a_{*}<0.7$.

Other models, such as that from Ebisawa et 
al. (1994), which assume $a_{*}=0$ are also incompatible with the mass and 
distance of G04.  These models suggest that the mass would need to be at least
$16~M_{\odot}$ or the distance would need to be less than $2.65~{\rm kpc}$ in
order to measure an inner radius consistent with $a_{*}=0$.  An additional way
to retrieve a retrograde spin would be to lower $f_{\rm col}$ to ~1.0, which is
suggested by our measured value existing near the hard limit of 1.5.  
If we allow $f_{\rm col}$ to be free, we find that its best fit is 
$f_{\rm col}=1.17^{+0.35}_{-0.28}$, but the spin no longer has any 
constraint.  We choose to use our range of $f_{\rm col}$ because these are the 
range of values expected for the range of luminosities in our sample (Shimura 
\& Takahara 1995).  The Ebisawa et al. (1994) model can also 
be fixed by increasing $r_{in}$, implying $a_{*}<0$.

To examine how $a_{*}$ is affected by the hard component of the spectrum, we 
tried using {\it powerlaw} instead of {\it simpl} to model the hard component.
This represents a less physical model than simpl, but it produces a check on 
our result.  For these fits $\Gamma$ and the normalization were allowed to be 
free and independant between spectra.  When this model is fitted to the data, 
we find the spin to be $a_{*}=-0.46_{-0.23}^{+0.02}$, consistent with the 
result obtained using {\it simpl}.  

\begin{figure}[tb]
\includegraphics[scale=0.45]{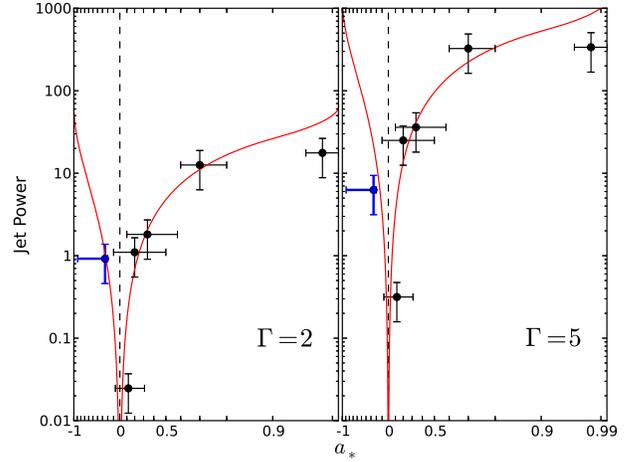}
\caption{\footnotesize Comparison of our measured spin and the calculated jet 
power (blue) with the best fit empirical model from Narayan \& McClintock 
(2012), and with H1743-322 (added in Steiner, McClintock \& Narayan 2013), 
having $\Gamma=2$ (left) or $\Gamma=5$ (right).}
  \vspace{0cm}
  \label{fig:NM12}
\end{figure}

It is also interesting to consider how our result fits into the context of
jet production, since very few retrograde spins have been observed, making GS 
1124-683 an interesting test of current empirical models.  One such model was
suggested by Narayan \& McClintock (2012), which suggests that the scaled jet
power is proportional to the black hole spin.  This model was used by Steiner, 
McClintock \& Narayan (2013) to predict the spins of 6 black holes including
GS 1124-683.  In their analysis, they only considered prograde spins, as no 
retrograde spins had been observed using the CF method at that time.  To 
calculate jet power, we used their prescription:
\begin{equation}
P_{\rm jet}=\left(\frac{\nu}{5~{GHz}}\right)~\left(\frac{S_{\nu,0}^{tot}}{Jy}\right)~\left(\frac{D}{\rm kpc}\right)^2~\left(\frac{M}{M_{\odot}}\right)^{-1}
\end{equation}

\noindent where $S_{\nu,0}^{\rm tot}$ is the beaming corrected flux.
\begin{equation}
S_{\nu,0}^{tot}=S_{\nu,obs}\times(\Gamma[1-\beta\cos{i}])^{3-\alpha}
\end{equation}

$\Gamma$ is the lorentz factor, which we assumed to be 2 (to compare to their 
derived relationship for $\Gamma=2$.  It should be noted that $\Gamma$ is 
likely not the same for all jet sources.), $\alpha$ is the radio spectral 
index, which is 0.5-0.6 for GS 1124-683 (Ball et al. 1995).  $\beta$ follows
from $\Gamma$, and $i$ is the inclination of the system, for which we used 
$54^{\circ}$ (G04), since we used the same value to measure the spin.
Using the maximum radio flux suggested by Ball et al. (1995) of 
$\approx0.2~{\rm Jy}$, the scaled jet power is then $\approx0.92$ in 
natural units.  Arbitrarily assuming an error in the radio flux of a factor of 
$\approx0.5$ (following the methodology of Narayan \& McClintock 2012), and 
using our determination of spin found from our steppar run, we find that our 
spin measurement is consistent with their best fit model (see 
Figure~\ref{fig:NM12}) which predicts $P_{\rm jet}=1.08^{+0.69}_{-0.43}$ for 
$a_{*}=-0.25$, It is different from the determination of the spin in Steiner, 
McClintock, \& Narayan (2013) only because they had assumed that the spin would
be prograde, and because they used values for mass and inclination different 
from the measured mass and inclination.  We note that a full consideration of 
the current data does not find strong evidence that spin powers jets, with 
$\dot{M}$ or $|{\textbf B}|$ potentially acting as a throttle (King et al. 
2013).

\section{conclusions}

 \noindent (1) For the most recent determinations of mass and distance to GS 
1124-683, the spin is most likely $-0.25^{+0.05}_{-0.64}$.  There is
an upper limit to the spin of -0.15 ($6\sigma$ level).  This result is 
independant of the model used to fit the hard component.

\noindent (2) Keeping the distance held within the constraints from G04, the 
minimum mass for GS 1124-683 where we can derive a 
prograde spin is $M=15.25~M_{\odot}$.  

\noindent (3) Keeping the mass held within the constraints from G04, the 
maximum distance from which we can potentially resolve a prograde 
spin is $d=4.5~{\rm kpc}$. 

\noindent (4) If we require the color correction factor $f_{\rm col}$ to be 
fixed at 1.7 for all spectra, the spin becomes pegged at the hard limit of -1. 
The upper limit necessary to avoid this is $f_{\rm col}=1.67$.

\noindent (5) GS 1124-683 agrees with the empirically derived 
relationship between black hole spin and jet power from Narayan \& McClintock
(2012) with $\Gamma=2$, though there are many caveats and assumptions.

\medskip

\end{document}